\documentclass[a4paper]{article}
\usepackage[comma,numbers,square,sort&compress]{natbib}
\newtheorem{lem}{Lemma}
\newtheorem{thm}{Theorem}

\newtheorem{rem}{Remark}
\newtheorem{ass}{Assumption}
\newtheorem{cor}{Corollary}

\usepackage{graphicx,amsmath,amssymb}
\usepackage{lineno,cases}
\usepackage{subfigure,pgf,tikz,pst-node}
\usetikzlibrary{arrows,shapes,automata,positioning,fit}

\newcommand{\pb}{\noindent\textbf{Proof.}: }
\newcommand{\pe}{\hfill\rule{4pt}{8pt}}

\def\rm{\mathrm}
\newcommand{\lm}{\lambda}
\newcommand{\e}{\varepsilon}
\newcommand{\R}{\mathbb{R}}
\newcommand{\w}{\omega}
\renewcommand{\top}{{\mathrm{T}}}

\begin{document}
	
	\title{Distributed Coordination for a Class of Nonlinear Multi-agent Systems with Regulation Constraints}
	
	\author{Yutao Tang, and Peng Yi		
		\thanks{Y. Tang is with the School of Automation, Beijing University of Posts and Telecommunications, Beijing, China (e-mail: yttang@bupt.edu.cn); P. Yi is with the Department of Electrical and Computer Engineering, University of Toronto, Toronto, Canada (email: peng.yi@utoronto.ca).}
	}
	
	\date{}
	
	\maketitle
	
	{\noindent\bf Abstract}: In this paper, a multi-agent coordination problem with steady-state regulation constraints is investigated for a class of nonlinear systems. Unlike existing leader-following coordination formulations, the reference signal is not given by a dynamic autonomous leader but determined as the optimal solution of a distributed optimization problem. Furthermore, we consider a global constraint having noisy data observations for the optimization problem, which implies that reference signal is not trivially available with existing optimization algorithms. To handle those challenges, we present a passivity-based analysis and design approach by using only local objective function, local data observation and exchanged information from their neighbors. The proposed distributed algorithms are shown to achieve the optimal steady-state regulation by rejecting the unknown observation disturbances for passive nonlinear agents, which are persuasive in various practical problems.  Applications and simulation examples are then given to verify the effectiveness of our design.
	\vskip 1.5ex

\maketitle

\section{Introduction}
Over the last decade, there has been an increasing percentage of literature concerning the coordination problem of multi-agent systems  due to its wide applications in engineering systems. e.g., multi-robot system and wireless network (see \cite{olfati2007consensus, renwei2011springer, park2015consensus} and references therein). As one fundamental problem of this topic, leader-following coordination has been widely studied (\cite{ni2010leader,wang2010distributed, tang2014leader, su2014unknown, tang2015distributed}). In this problem, a (virtual) leader is often set up to generate the reference signals for each agent to follow, while this leader is usually given as a known dynamic system with possible unknown states. Then the main task is to determine the agents' controllers, which should only utilize local information, such that the resultant state or output trajectories of the agents can track the reference signal generated by the leader. This leader-following formulation has been effectively embodied into various effective algorithms for practical engineering problems, e.g, formation control and attitude synchronization (\cite{renwei2011springer,thunberg14auto}).

Here, we follow this line but consider a particular case when the reference signal is not generated as the trajectory of an autonomous leader, but as the unknown optimal solution of distributed optimization problems.  This type of problems arises naturally from many practical applications. For example, in a source seeking problem, we aim to control one or more agents with nonlinear dynamics to seek the extremum of some unknown signal field based on local signal measurements. Thus the reference (although a constant) is neither available in advance nor can be generated by an autonomous leader without real-time measurements and computations. Many other practical engineering applications have a similar feature that the reference signal is a (time-varying) maximum or minimum of some performance function, e.g., the design of anti-lock braking systems (\cite{zhang2011extremum}), optimal rendezvous of unmanned aerial vehicles (\cite{qiu2016distributed}). Moreover, practically each agent can only access noisy local observations/measurements. How to solve these optimal regulation problems over multi-agent systems with data locality and impurity is important and challenging. 

Furthermore, the so-called distributed optimization becomes more and more popular with their broad applications in various fields such as intersection computation (\cite{shi2013reaching}) and smart grids (\cite{bose2015equivalent}). In this problem, each agent only knows its own local cost function, and may only know its locally observable data. The agents aim to cooperatively achieve a consensus on their states and optimize the sum of all local cost functions. Both discrete-time and continuous-time gradient-based optimization algorithms were proposed (\cite{nedic2009distributed, yuan2015gradient, lin2016distributed,tang2016icca}).  In fact, we can regard this problem as a leader-following problem when the agents are single integrators and the reference signal is determined as the optimal solution. We may wonder its solvability when agents are of high-order physical processes, e.g., motion of mobile sensors and the dynamics of the inventory system. Since the decision variables are determined as outputs of high-order nonlinear dynamic systems that essentially cannot be treated as single integrators, the solvability of those distributed optimization problems over physical dynamics can be much more challenging than the conventional cases.

Based on  these observations, we aim to formulate and investigate a distributed coordination problem over high-order nonlinear multi-agent systems where the reference signal is determined as the optimal solution to a constrained optimization problem. Specifically, we consider a class of passive nonlinear agents, and the output variables are regulated according to a distributed resource allocation problem. The goal is then to regulate these outputs to the optimal solution of the associated optimization problem in a distributed way.

In fact, very few optimization results have been obtained on this topic with agents with high-order dynamics, though there are plenty of conclusions obtained for integrator-type agents as mentioned above. Since practical systems are hardly described by single integrators, we have to take high-order dynamics into consideration. For example, for double integrators, the authors in \cite{zhang2014distributed} proposed a distributed optimization algorithm with an integral control idea and a similar design was employed for Lagrangian agents (\cite{Deng}). Distributed optimization with input disturbances was also considered in \cite{wang2016distributed} by an internal-model approach for a class of nonlinear minimum phase agents. An important engineering problem related to this topic is the economic dispatch in power systems. While economic dispatch can be formulated and solved as a distributed optimization problem (\cite{cherukuri2015distributed, yi2016initialization}), frequency dynamics were taken to consideration in \cite{trip2016internal} and { an optimization requirement on the steady-state inputs should be satisfied. } An internal-model based controller was proposed to solve the optimal frequency regulation problem in power grids under unknown and possible time-varying load changes.  Generally, resource allocation over nonlinear multi-agent systems is still far from being solved.

In view of the aforementioned results, the main contributions of this paper are at least two-fold:

	(i) A general multi-agent coordination problem with regulation constraints is formulated for a class of nonlinear agents.  Since he steady-state of each agent can not be obtained in a centralized way and can only be approached asymptotically  through a distributed computation within the networks, it can be taken as distributed extensions of conventional steady-state regulation problems (\cite{jokic2009constrained, zhang2011extremum}). Distributed algorithms are proposed to solve this problem by using only local data and exchanged information from their neighbors, along with both the asymptotic and exponential convergence results, while only local and asymptotic results were considered in \cite{jokic2009constrained}.
	
	(ii) A passivity-based approach is adopted to distributedly solve resource allocation problem over high-order nonlinear multi-agent systems. When agents are single integrators without observation disturbances, this problem reduces to the conventional distributed resource allocation problem (\cite{cherukuri2015distributed, yi2016initialization}). Note that the decision variables are outputs of physical plants with high-order nonlinear dynamics, and hence this problem is much more challenging than traditional cases. This work shows that the passivity-based approach could provide a fresh design methodology to solve this kind of problems for high-order dynamic agents, including distributed inventory control (\cite{raafat1991survey}) and average consensus problem (\cite{renwei2011springer, rezaee2015average}).
	
	Additionally, unknown data observation disturbances are considered in our formulation and then rejected by an observer-based compensator. This approach is different from existing internal model-based designs (\cite{wang2016distributed, trip2016internal}), and may provide new perspectives and approaches to deal with this kind of problems.

The paper is organized as follows. Preliminaries about convex analysis, graph theory and passivity are given in Section \ref{sec:pre}. Then the problem of distributed coordination with regulation constraints is formulated for a class of nonlinear multi-agent systems in Section \ref{sec:formulation}. Main results are presented and proved in Section \ref{sec:main} along with the proposed gradient-based controls. Following that, three applications are given with discussions in Section \ref{sec:discussion} to illustrate the applicability and effectiveness of the proposed algorithm. Finally, concluding remarks are given in Section \ref{sec:con}.

\textsl{Notations:} Let $\R^n$ be the $n$-dimensional Euclidean space. For a vector $x$, $||x||$ denotes its Euclidean norm. ${\bf 1}_N$ (and ${\bf 0}_N$) denotes an $N$-dimensional all-one (and all-zero) column vector. $\mbox{col}(a_1,\,{\dots},\,a_n) = [a_1^\top,\,{\dots},\,a_n^\top]^\top$ for column vectors $a_i\; (i=1,\,{\dots},\,n)$. $r_N=\frac{1}{\sqrt{N}}{\bf 1}_N$, and $R_N\in \mathbb{R}^{N \times (N-1)}$ satisfying $R_N^\top r_N={\bf 0}_N$, $R_N^\top R_N=I_{N-1}$ and $R_N R_N^\top=I_{N}-r_N r_N^\top$.

\section{Preliminaries}\label{sec:pre}

In this section, preliminaries are given about convex analysis (\cite{bertsekas2009convex}), graph theory (\cite{mesbahi2010graph}) and system passivity (\cite{khalil2002nonlinear}).

\subsection{Convex analysis}

A function $f(\cdot)\colon \R^N \rightarrow \R $ is said to be convex if for any $0\leq a \leq 1$,
$$ f(a\zeta_1+(1-a)\zeta_2)\leq af(\zeta_1)+(1-a)f(\zeta_2), ~ \forall \; \zeta_1,\zeta_2 \in \R^N.$$
A differentiable function $f$ is convex over $\R^N$ if
\begin{equation}\label{eq:def-convex}
f(\zeta_1)-f(\zeta_2)\geq \nabla f(\zeta_2)^T(\zeta_1 -\zeta_2),~ \forall \; \zeta_1,\zeta_2 \in \mathbb{R}^N,
\end{equation}
and $f$ is strictly convex over $\mathbb{R}^N$ if the above inequality is strict whenever $\zeta_1 \neq \zeta_2$,
and $f$ is $\omega$-strongly convex ($\omega >0$) over $\mathbb{R}^N$ if $\forall ~ \zeta_1, \zeta_2 \in \mathbb{R}^N$,
\begin{eqnarray}\label{eq:def-sc}
(\nabla f(\zeta_1)-\nabla f(\zeta_2))^T(\zeta_1 -\zeta_2)\geq \omega \|\zeta_1 -\zeta_2\|^2.
\end{eqnarray}
A function $f: \mathbb{R}^N \rightarrow \mathbb{R}^N$ is Lipschitz
with constant $M>0$, or simply $M$-Lipschitz, if
$$
\|f(\zeta_1)-f(\zeta_2)\|\leq M \|\zeta_1-\zeta_2\|, ~ \forall ~ \zeta_1, \zeta_2 \in \mathbb{R}^N.
$$

\subsection{Graph theory}

A weighted undirected graph is described by $\mathcal {G}=(\mathcal {N}, \mathcal {E}, \mathcal{A})$ with the node set $\mathcal{N}=\{1,{\dots},N\}$ and the edge set $\mathcal {E}$. $(i,\,j)\in \mathcal{E}$ denotes an edge between nodes $i$ and $j$. The weighted adjacency matrix $\mathcal{A}=[a_{ij}]\in
\mathbb{R}^{N\times N}$ is defined by $a_{ii}=0$ and
$a_{ij}=a_{ji}\geq0$ ($a_{ij}>0$ if and only if there is an edge
between node $i$ and node $j$). The neighbor set of node $i$ is
defined as $\mathcal{N}_i=\{j: (j,i)\in \mathcal {E} \}$ for
$i=1,\,...\,,n$.  A path in graph $\mathcal{G}$ is an alternating
sequence $i_{1}e_{1}i_{2}e_{2}{\dots}e_{k-1}i_{k}$ of nodes $i_{l}$
and edges $e_{m}=(i_{m},i_{m+1}) \in\mathcal {E}$ for
$l=1,2,{\dots},k$. If there is a path between any two vertexes of a
graph $\mathcal{G}$, then the graph is said to be connected.  The
Laplacian $L=[l_{ij}]\in \mathbb{R}^{N\times N}$ of graph
$\mathcal{G}$ is defined as $l_{ii}=\sum_{j\neq i}a_{ij}$ and
$l_{ij}=-a_{ij} (j\neq i)$, which is thus symmetric.  Denote the
eigenvalues of Laplacian matrix $L$ associated with an undirected
graph $\mathcal{G}$ as $\lm_{1}\,\leq \,\dots\,\leq
\,\lm_{N}$.



\subsection{System passivity}
Passivity, due to its explicit physical meaning and simplicity to manipulate, has been extensively discussed in controlling various practical engineering systems (see \cite{khalil2002nonlinear} and references therein).  Usually, only the case when the equilibrium point is zero is investigated. However, when desirable regulation points are specified as solutions to optimization problems, we do not know the optimal point beforehand and have to go back to the general case with  non-zero equilibrium points, which has been named as incremental passivity in some literature (\cite{wen2004unifying, jayawardhana2007passivity, pavlov2008incremental, stegink2017unifying}).

Consider a dynamic system of the following form:
\begin{align}\label{sys:def}
\dot{x}=g(x,\,u),\quad  y=h(x),\quad x\in \mathbb{R}^n,\, u,\,y\in\mathbb{R}^p.
\end{align}
Let $\mathbb{E}=\{(x^*,\,u^*)\mid g(x^*,\,u^*)=0\}$ be the equilibrium points of this system. System \eqref{sys:def} is said to be passive with respect to (w.r.t.) $(x^*,\,u^*)$ if there exist two functions $\alpha_1(\cdot),\, \alpha_2(\cdot) \in \mathcal{K}_\infty$ and a continuously differentiable storage function $V(x,\,x^*)$ satisfying that:
\begin{itemize}\label{def:condition}
	\item[i)] $\alpha_1(||x-x^*||)\leq V(x,\,x^*)\leq \alpha_2(||x-x^*||)$
	\item[ii)]$\dot{V}\leq (y-y^*)^\top (u-u^*)$ with $y^*=h(x^*)$
\end{itemize}
where $\dot{V}$ is short for $\dot{V}(x,\,x^*)\triangleq \frac{\partial V}{\partial x}(x,\,x^*)\dot{x}$.

When the second condition is strengthened as
\begin{align*}
\dot{V}\leq -\alpha_3(||x-x^*||)+(y-y^*)^\top (u-u^*)
\end{align*}
for some function $\alpha_3(\cdot)\in \mathcal{K}$, this system is said to be strictly passive w.r.t. $(x^*,\,u^*)$. If $\alpha_1(\cdot),\, \alpha_2(\cdot),\, \alpha_3(\cdot)$ can be taken as quadratic functions, this system is said to be exponentially passive w.r.t. $(x^*,\,u^*)$.   For consistence, a memoryless function $\phi\colon\mathbb{D}\subset\mathbb{R}^n\to \mathbb{R}^n$ is passive  w.r.t. $y^*\in\mathbb{R}^n$ if it satisfies
\begin{align}
(y-y^*)^\top (\phi(y)-\phi(y^*))\geq 0,\quad \forall y\in \mathbb{D}.
\end{align}
When the equality occurs only if $y=y^*$, we say it is strictly passive w.r.t. $y^*$. Additionally, when there exists a constant $\gamma>0$ such that
\begin{align*}
(y-y^*)^\top (\phi(y)-\phi(y^*))\geq\gamma||y-y^*||^2,\quad \forall y\in \mathbb{D},
\end{align*}
it is exponentially passive with modulus $\gamma$ w.r.t. $y^*$.

In this paper,  we only  consider the single-input single-output case, i.e., $p=1$, while the following arguments hold for the multi-input multi-output case with $p>1$ as well.   For a given passive system w.r.t. $(x^*,\,u^*)$, this equilibrium is said to be assignable if there exists a passive $\phi(\cdot)$ satisfying $\phi(y^*)+u^*=0$. The following lemma presents an important approach to stabilize nonlinear passive systems.

\begin{lem}\label{lem:monotone}
	Suppose system \eqref{sys:def} is passive w.r.t. an assignable equilibrium $(x^*,\,u^*)$, then, $x=x^*$ is Lyapunov stable under $u=-\phi(y)$. If $\phi(\cdot)$ is strictly passive w.r.t. $y^*$, we have $y(t)\to y^*$ as $t$ goes to infinity. Moreover, the trajectory $x$ converges to $x^*$ exponentially fast if system \eqref{sys:def} is exponentially passive w.r.t. $(x^*,\,u^*)$.
\end{lem}
\pb  The proof follows standard Lyapunov arguments.  In fact, when the system is passive w.r.t. $(x^*,\, u^*)$, we have a continuously differentiable storage function $V(x,\,x^*)$ and functions $\alpha_1,\, \alpha_2 \in \mathcal{K}_\infty$ satisfying: $\alpha_1(||x-x^*||)\leq V(x,\,x^*)\leq \alpha_2(||x-x^*||)$ and $\dot{V}\leq (y-y^*)^\top (u-u^*)$ with $y^*=h(x^*)$.
	
	By taking $u=-\phi(y)$, we have $\dot{V}\leq - ( y-y^*)^\top (\phi(y)-\phi(y^*))$. From the strict passivity of $\phi(\cdot)$, it implies
	$\dot{V}\leq 0$ and the equation happens only if $y=y^*$. {  By LaSalle's invariance principle (\cite{khalil2002nonlinear}) and the smoothness of $h(\cdot)$, we can obtain $x(t)$ converges to the largest invariant set contained in $\{x\in \R^n | h(x)=h(x^*)\}$ as $t\to +\infty$, which implies $y(t)\to y^*$ as $t$ goes to infinity.}
	
	When this system is exponentially passive, we further have $k_1||x-x^*||^2\leq V(x,\,x^*)\leq k_2||x-x^*||^2$ and $\dot{V}\leq -k_3 ||x-x^*||^2$, which imply the exponential convergence of $x$ w.r.t. $x^*$.
\pe 

By this lemma, the asymptotic regulation of $y$ to $y^*$ is transformed into a problem that to find a passive function $\phi(y)$ satisfying $\phi(y^*)+u^*=0$. For a special case when $x^*=0,\, u^*=0$, we need to find a monotone  (passive) function $\phi(\cdot)$ vanishing at the origin, which is consistent with existing results. It is well-known that every continuous (strictly, strongly) convex function has its derivative (including gradients as its special cases) as an associated (strictly, exponentially) passive function w.r.t. the minimum point (\cite{bauschke2011convex}). This observation will play a key motivation for our gradient-based algorithm design when we do not have the direct information of $x^*$ and hence $y^*$. Although we consider gradient-based designs, various functions (not limited to gradients) can be employed in a passivity-based design for different problems, e.g. skew-symmetric linear operators and saddle-point operators (\cite{bauschke2011convex, gadjov2017passivity}).


\section{Problem Formulation}\label{sec:formulation}

Consider $N$ agents with dynamics of the form:
\begin{align}\label{sys:agent}
\dot{x}_i=g_i(x_i,\,u_i),\quad y_i=h_i(x_i),\quad i=1,\,\dots,\,N
\end{align}
with state variable $x_i\in \mathbb{R}^{n_i}$, control input $u_i\in\mathbb{R}$, and output $y_i\in \mathbb{R}$. The functions $g_i(\cdot)$ and $h_i(\cdot)$ are assumed to be smooth.  

Along with node dynamics, the $i$-th agent has a local cost function $f_i(y_i)$. For this multi-agent system, we associate it an optimization problem with coupled constraints as follows.
\begin{align}\label{opt:ra}
\begin{split}
\mbox{minimize}&\quad  \quad f(y)\triangleq\sum\nolimits_{i=1}^Nf_i(y_i)\\
\mbox{subject to}&\quad   ~~\sum\nolimits_{i=1}^N y_i=\sum\nolimits_{i=1}^N d_i^0
\end{split}
\end{align}
where $d_i^0$ can only be obtained by agent $i$ by local measurements. {  This optimization problem is often called resource allocation (\cite{bertsekas1998network}) and many practical applications can be formulated as the above, e.g. economic dispatch in power systems (\cite{yi2016initialization}), flow control in networks (\cite{bertsekas1998network}). We can regard $y_i$ as the amount of resource located at node $i$ and interpret $-f_i(y_i)$ as the local (concave) utility function. Coupled with the physical agents, we aim to design distributed controllers such that the outputs of these agents asymptotically solve the optimization problem \eqref{opt:ra}. This implies the above optimization problem is a requirement on the steady-state of this multi-agent system. In other words, the controller should regulate the systems' outputs such that the equality constraint is satisfied and optimal performance is achieved, both in an asymptotic manner.}

Moreover, we are interested in distributed algorithms without setting up a centralized working station which might be expensive and inhibitive in some circumstances.  Namely, we aim to find a distributed protocol using only local objective function, constraint related observation, and exchanged information from their neighbors to drive the outputs of agents to reach an allocation that maximizes the total utility $-\sum_{i=1}^Nf_i(y_i)$. For this purpose, an undirected graph $\mathcal{G}$ is employed to describe the information sharing relationships among those nodes represented by $\mathcal{N}=\{1,\,\dots,\, N\}$. If node $i$ and $j$ can exchange information with each other, then there is an edge $(i,j)$ in the graph $\mathcal{G}$, i.e., $a_{ij}=a_{ji}>0$.

Moreover, we assume agent $i$ can only get a polluted observation $d_i(t)\triangleq d_i^0+d_i^\e(t)$ of $d_i^0$ by an imperfect sensor, where the disturbance is assumed consisting of $k_i$ sinusoidal signals with distinct but known frequencies: $\omega_{i1},\dots,\,\omega_{ik_i}$. In fact, this type of disturbances can produce a fair approximation of any bounded periodic disturbance signal by summing up the dominated harmonics in its Fourier series expansion and has been used as typical nontrivial disturbances in the control literature (\cite{huang2004nonlinear, chen1995linear}).

We then formulate the distributed coordination problem for nonlinear multi-agent systems with steady-state regulation constraints as follows. {\em Given the graph $\mathcal{G}$, cost function $f_i(\cdot)$ and dynamic plant \eqref{sys:agent}, find a distributed control $u_i$ for agent $i$, which only depends on its own local data and exchanged information from its neighbors, such that the trajectories of agents are bounded and satisfy \[\lim_{t\to +\infty} y_i= y^*_i,\quad  \mbox{for }\; i=1,\,\ldots,\,N \] where $\mbox{col}(y_1^*,\,\dots,\, y_N^*)$ is the optimal solution of \eqref{opt:ra}.

\begin{rem}
	Unlike existing leader-following coordination problems in multi-agent systems (\cite{ni2010leader, renwei2011springer, su2014unknown}), the reference signal can not be modeled as an autonomous leader. In fact, the desired steady states can only be obtained by a distributed cooperation and computation. {  On the other hand, when $f_i(y_i)=\frac{1}{2}y_i^2$ and $d^0_i=y_i(0)$, the optimal solution of \eqref{opt:ra} is $\frac{\sum_{i=1}^N y_i(0)}{N}{\bf 1}_N$. Our formulation provides another way to solve the well-known average consensus problem for nonlinear dynamic agents (\cite{xiao2004fast, renwei2011springer, rezaee2015average}).}
\end{rem}


Our formulation can be taken as a constrained optimization problem subject to noisy constraint data observations. Compared with the traditional stochastic or the worst-case formulation in robust optimization,  the observation perturbations/uncertainties  here are modeled as structured but unknown ones, which can be deemed as a balance on available information of perturbations between its nominal version and its stochastic/worst-case version.  

To achieve a coordination among these agents, some technical assumptions are needed.

\begin{ass}\label{ass:function}
	For $i=1,\,\dots,\, N$, the function $f_i\colon\R\to\R$ is convex, twice continuously differentiable with bounded Hessian, i.e., there exist $0 < \underline{h}_i  \leq \overline h_i <\infty$  such
	that, for all $i$:
	$$\underline{h}_i 	 \leq \nabla^2 f_i(s) \leq \overline h_i ,\quad \forall s \in  \R.$$
\end{ass}

\begin{ass}\label{ass:graph}
	The communication graph $\mathcal{G}$ is connected.
\end{ass}

The assumptions have been widely used in many publications (\cite{mesbahi2010graph, kia2015distributed, tang2015distributed}). {  Assumption \ref{ass:function} is made to guarantee the solvability of this optimization problem. In fact, it implies the Lipschitz continuity of  $\nabla f_i$ and strong convexity of $f_i$. Then, the optimization problem \eqref{opt:ra} is solvable and has a unique solution $ y^*=\mbox{col}(y_1^*,\,\dots, \, y_N^*)$ by Proposition 3.2.1 in \cite{bertsekas2009convex}.}  It is well-known (\cite{mesbahi2010graph}) that under Assumption \ref{ass:graph}, the associated Laplacian $L$ of this graph is symmetric with rank $N-1$, and its null space is spanned by ${\bf 1}_N$.

Another technical assumption is made as follows.
\begin{ass}\label{ass:regulator}
	For any $i\in \{1,\,\dots,\,N\}$ and constant output $r$, there exist two unique smooth functions ${\bf x}_i(\cdot)$ and ${\bf u}_i(\cdot)$ satisfying
	\begin{align}
	g_i({\bf x}_i(r),\,{\bf u}_i(r))=0, \quad r=h_i({\bf x}_i(r)).
	\end{align}
	Furthermore, the function ${\bf u}_i(\cdot)$ is $M_i$-Lipschitz at its arguments on the concerned set.  
\end{ass}

Note that our problem is essentially an asymptotic regulation problem where the reference is determined by the optimization problem \eqref{opt:ra}, thereby, Assumption \ref{ass:regulator} can be understood as the existence of the solution to regulator equations in the terminology of output regulation. The Lipschitzness of ${\bf u}_i(\cdot)$ is only made for technical analysis, and it naturally holds when the concerned sets are compact. Similar assumptions can be found in \cite{khalil2002nonlinear, huang2004nonlinear}.}

Denoting ${x}_i^*={\bf x}_i(y^*_i)$ and ${u}_i^*={\bf u}_i(y^*_i)$, we then focus on a class of passive nonlinear dynamic systems as follows.
\begin{ass}\label{ass:passivity}
	For $i=1,\,\dots,\, N$, the dynamics \eqref{sys:agent} is passive w.r.t. $(y^*_i,\, x^*_i,\, u^*_i)$, i.e., there exists a continuously differentiable storage function $V_i(x_i,\,x^*_i)$ satisfying  $\alpha_{i1}(||x_i-x^*_i||)\leq V_i(x_i,\,x^*_i)\leq \alpha_{i2}(||x_i-x_i^*||)$ for two $\mathcal{K}_\infty$ functions $\alpha_{i1}(\cdot)$ and $\alpha_{i2}(\cdot)$ such that
	$$\dot{V}_i\leq (y_i-y_i^*)^\top (u_i-u_i^*).$$
\end{ass}

As mentioned before, this condition is an extended version of classical passivity property with respect to nonzero equilibria (by nonzero inputs).  In fact,  it has been termed as {\bf incremental passivity} property in many publications (\cite{wen2004unifying, jayawardhana2007passivity, pavlov2008incremental, stegink2017unifying}), and a large class of typical systems falls into this class perhaps after an inner feedback passivation loop. Notice that Assumption \ref{ass:passivity} only concerns with the dynamics of the nonlinear agents and imposes no restrictions on the optimization problem. 

In the following section, we solve the distributed coordination problem with regulation constraints by passivity-based arguments.

\section{Main Results} \label{sec:main}
In this section, we propose a distributed algorithm to solve the distributed coordination problem determined by \eqref{sys:agent} and \eqref{opt:ra}, and then prove its stability via passivity techniques.


	%
	%
	
	First, we split the control efforts into two parts $u_i=u_i^1+u_i^2$, where $u_i^1$ is designed for distributed optimization with disturbance rejection and $u_i^2$ for asymptotic steady-state regulation. 
	
	Inspired by those works in \cite{bertsekas1998network, kia2015distributed,  yi2016initialization}, the optimization problem \eqref{opt:ra} can be rewritten as a monotropic programming problem and solved by a primal-dual approach if $d^0$ is known. Since there are observation disturbances for agent $i$ in $d^0_i$, an observer-based approach is employed to estimate $d^0_i$ while rejecting those disturbances $d_i^\e$.  Then, the first part of our control is given as follows:
	\begin{align}\label{ctr:robust-1}
	\begin{split}
	u_i^1&=-\gamma \nabla f_i(y_i)+\lm_i\\
	\dot{\lm}_i&=-\lm^v_i-z^v_i+d_i-y_i-D_i^\epsilon \eta_i\\
	\dot{\eta}_i&=(S_i-L_iD_i)\eta_i+L_id_i\\
	\dot{z}_i&=\lm^v_i,\quad i=1,\,\dots,\,N
	\end{split}
	\end{align}
	where $S_i=\mbox{diag}\left(0, \, \begin{bmatrix}
	0&\omega_{i1}\\
	-\omega_{i1}&0
	\end{bmatrix},\,\dots,\,\begin{bmatrix}
	0&\omega_{ik_i}\\
	-\omega_{ik_i}&0
	\end{bmatrix}\right)$,\, $D_i=[1, \underbrace{1,0,\, \dots,\, 1,\,0}_{2k_i}]$,\, $D_i^\epsilon=[\underbrace{1,0,\, \dots,\, 1,\,0}_{2k_i}]$, $\lm^v_i\triangleq \sum\nolimits_{j=1}^N a_{ij}(\lm_i-\lm_j)$, $z^v_i\triangleq \sum\nolimits_{j=1}^N a_{ij} (z_i-z_j)$,  and  $\gamma\in \R, \,L_i\in \R^{(2k_i+1)\times 1}$ will be determined later. Here, the parameter $\gamma$ can be regarded as a proportional gain to correct the regulation error, and the $\eta_i$-subsystem is a local observer to estimate $d_i(t)$ for disturbance rejection. 

The following lemma guarantees the selection of gain matrix $L_i$ such that $S_i-L_iD_i$ is Hurwitz and the effectiveness of our algorithm in disturbance rejection.
\begin{lem}\label{lem:observability}
	The pair $(S_i\,\ D_i)$ is observable.
\end{lem}
\pb We consider the rank of $[S_i^\top-\lambda I_{2k_i+1}, D_i^\top]$. When $\lambda\notin \{0,\, \w_1,\,\dots,\,\w_{k_i}\}$, this is obvious.
	
	Suppose $\lambda=0$, the matrix $[S_i^\top,\,D_i^\top]$ can be partitioned as follows:
	\begin{align*}
	\begin{bmatrix}
	0 & {\bf 0}_{2k_i}^\top &1\\
	{\bf 0}_{2k_i}&\hat S_i^\top&{D^\e_i}^\top
	\end{bmatrix}
	\end{align*}
	where $\hat S_i=\mbox{diag}\left(\begin{bmatrix}
	0&\omega_{i1}\\
	-\omega_{i1}&0
	\end{bmatrix},\,\dots,\,\begin{bmatrix}
	0&\omega_{ik_i}\\
	-\omega_{ik_i}&0
	\end{bmatrix}\right)$. Since the rank of $\hat S_i$ is $2k_i$, ${\rm rank}[S_i^\top,\,D_i^\top]$ is then $2k_i+1$.
	
	Suppose $\lambda=\w_{ij}$, without loss of generalities $j=k_i$ and partition $[S_i^\top-\w_{i k_i},\,D_i^\top]$ in the following form:
	$$\begin{bmatrix}
	-\lambda 					& {\bf 0}_{2k_i-2}^\top 					&{\bf 0}_2^\top															&1\\
	{\bf 0}_{2k_i-2}			&\tilde S_i^\top-\lambda I_{2k_i-2}				& [{\bf 0}_{2k_i-2}~~{\bf 0}_{2k_i-2}]								&\tilde {D^\e_i}^\top\\
	{\bf 0}_2					&\begin{bmatrix}{\bf 0}_{2k_i-2}^\top \\ {\bf 0}_{2k_i-2}^\top \end{bmatrix} 	& \begin{bmatrix}-\w_{iki}&-\w_{ik_i}\\\w_{ik_i} &-\w_{ik_i} \end{bmatrix}				&\begin{bmatrix}1\\0\end{bmatrix}
	\end{bmatrix}
	$$
	where $\tilde S_i=\mbox{diag}\left(\begin{bmatrix}
	0&\omega_{i1}\\
	-\omega_{i1}&0
	\end{bmatrix},\,\dots,\,\begin{bmatrix}
	0&\omega_{ik_i-1}\\
	-\omega_{ik_i-1}&0
	\end{bmatrix}\right)$ and $\tilde {D^\e_i}=[\underbrace{1,0,\, \dots,\, 1,\,0}_{2k_i-2}]$. It can be checked that the rank of $\begin{bmatrix}-\w_{iki}&-\w_{ik_i}&1 \\\w_{ik_i} &-\w_{ik_i}&0\end{bmatrix}$ is $2$ and the rest part has a rank $2k_i-1$, thus ${\rm rank}[S_i^\top-\w_{iw_{k_i}} I_{2k_i+1},\,D_i^\top]$ is then $2k_i+1$.
	
	To sum up, we obtain ${\rm rank}[S_i^\top-\lambda I_{2k_i+1},\,D_i^\top]=2k_i+1$ holds for any $\lambda$. By PHB-test (\cite{chen1995linear}), this implies the conclusion. \pe 

	For the second part of our control, we recall Assumption \ref{ass:regulator} and let $u^2_i={\bf u}_i(y_i)$. The whole control $u_i$ to achieve our designed goal is presented as follows:
	\begin{align}\label{ctrl:robust}
	\begin{split}
	u_i&=u_i^1+u_i^2=-\gamma \nabla f_i(y_i)+\lm_i+{\bf u}_i(y_i)\\
	\dot{\lm}_i&=-\lm^v_i-z^v_i+d_i-y_i-D_i^\epsilon \eta_i\\
	\dot{\eta}_i&=(S_i-L_iD_i)\eta_i+L_id_i\\
	\dot{z}_i&=\lm^v_i
	\end{split}
	\end{align}
	where the matrices $S_i,\, D_i,\, D_i^\epsilon,\, L_i$ are defined as above.

Under the information sharing constraints, {  the following lemma shows that at the equilibrium point $(\tilde x_i,\, \tilde \lm_i,\, \tilde z_i)$ of the closed-loop system composed of \eqref{sys:agent} and \eqref{ctrl:robust},} the associated output $\tilde y_i=h_i(\tilde x_i)$ actually solves the resource allocation problem \eqref{opt:ra} with disturbance rejection.
\begin{lem}\label{lem:opt-to-stability}
	Under Assumptions \ref{ass:function}--\ref{ass:regulator}, the equilibrium point of the closed-loop system composed by \eqref{sys:agent} and \eqref{ctrl:robust} satisfies the following conditions for some constant $\lm_0$:
	\begin{align}\label{eq:kkt}
	\begin{split}
	\nabla_{y_i} f_i(\tilde y_i)+\lm_0=0, ~ \sum_{i=1}^N \tilde  y_i=\sum\nolimits_{i=1}^N d_i^0, ~ i=1,\, \dots,\, N.
	\end{split}
	\end{align}
\end{lem}
\pb  {  The polluted observation $d_i$ can be rewritten into a form of $d_i(t)=A_{i0}+\sum_{j=1}^{k_i}A_{ij}\sin(\omega_{ij} t+\varphi_{ij})$, where $A_{i0}=d_i^0$, $A_{ij}$ and $\varphi_{ij}$ are unknown.  Then, by taking a proper state variable $\xi_i^d\in \R^{2k_i+1}$, we can put it into $\dot{\xi}^d_i=S_i \xi^d_i$ with $d_i(t)=D_i \xi_i^d$ and an initial condition $\xi^d_i(0)$ determined by $A_{ij}$ and $\varphi_{ij}$. 
		
		Letting $\overline d_i=\eta_i-d_i$ and recalling that $\dot{\eta}_i=(S_i-L_iD_i)\eta_i+L_id_i$, we have 
		\begin{align*}
		\dot{\overline d}_i&=\dot{\eta}_i-\dot{\xi}^d_i=(S_i-L_iD_i) \overline {d}_i.
		\end{align*}
	} 
	
	By the selection of $L_i$, the matrix $S_i-L_iD_i$ is Hurwitz and the trajectory of $\overline d_i(t)$ goes to 0 as time goes to infinity. Then the equilibrium point of the closed-loop system composed by \eqref{sys:agent} and \eqref{ctrl:robust} can be obtained by setting the derivatives of states to zero, i.e., for $i=1,\,\dots,\,N$,
	\[
	g_i(\tilde x_i,\, \tilde u_i)=0,~ -\tilde \lm^v_i-\tilde z^v_i+d_i^0-\tilde y_i=0,~ \tilde \lm^v_i=0.
	\]
	
	From $\lm^v_i=0$ and by Assumption \ref{ass:graph}, we have $\tilde \lm_1=\dots=\tilde \lm_N=\tilde \lm$ for some $\tilde \lm$. By summing up the second equation from $1$ to $N$, it follows $\sum_{i=1}^N \tilde  y_i=\sum\nolimits_{i=1}^N d_i^0$, where we use  ${\bf 1}^\top L={\bf 0}$.  By Assumption \ref{ass:regulator}, for given $\tilde y_i$ and by the uniqueness of ${\bf x}_i(\cdot),\, {\bf u}_i(\cdot)$, we have $-\gamma \nabla f_i(\tilde y_i)+\tilde\lm_i+{\bf u}_i(\tilde y_i)={\bf u}_i(\tilde y_i)$ and thus $\nabla f_i(\tilde y_i)-\frac{\tilde \lm}{\gamma}=0$. Let $\lm_0=\frac{\tilde \lm}{\gamma}$, the conclusion is thus complete.
\pe 

Since Assumption \ref{ass:function} implies that the optimization problem \eqref{opt:ra} has a unique solution, and thus $\tilde y_i=y^*_i, \, {\bf x}_i(y_i^*)={\bf x}_i(\tilde y_i), \,{\bf u}_i(y_i^*)={\bf u}_i(\tilde y_i)$ by Assumption \ref{ass:regulator}.Let $y_i^*, \,x_i^*, \,u_i^*$ without confusions represent $\tilde y_i,\, {\bf x}_i(\tilde y_i)$ and ${\bf u}_i(\tilde y_i)$ to save notations.

It is time to present our first main theorem.
\begin{thm}\label{thm:robust}
	Under Assumptions \ref{ass:function}--\ref{ass:passivity}, the distributed coordination problem with regulation constraints determined by \eqref{sys:agent} and \eqref{opt:ra} can be solved by the algorithm \eqref{ctrl:robust} with $\gamma > \max_{i}(\frac{1+M_i}{\underline{h}_i})$ and $L_i$ such that $S_i-L_iD_i$ is Hurwitz, i.e., $\lim_{t\to +\infty} y_i(t)= y^*_i$ for $i=1,\,\ldots,\,N$, where $\mathrm{col}(y_1^*,\,\dots,\, y_N^*)$ is the optimal solution of \eqref{opt:ra}. Moreover, if agent $i$ is $y^*_i$-observable, the closed-loop system is asymptotically stable at its equilibrium point.
\end{thm}
\pb  By Lemma \ref{lem:opt-to-stability}, we only have to show the stability and output convergence of the closed-loop system w.r.t. its equilibrium point.  
	
	Recalling the definition of passivity w.r.t. non-zero equilibria, the convergence part is trivial if a dynamic system is passive w.r.t. its equilibrium point. We next show the passivity of this closed-loop system with output $y=\mbox{col}(y_1,\,\dots,\,y_N)$ and a new control $\hat u \triangleq \mbox{col}(u_1-\lm_1,\,\dots,u_N-\lm_N)$ .
	
	In fact, since $\overline S_i=S_i-L_iD_i$ is Hurwitz, there exists a positive definite matrix $P_i\in \R^{(2k_i+1)\times (2k_i+1)}$, such that $\overline S_i^\top P_i+P_i\overline S_i=-I_{2k_i+1}$. We then consider a candidate storage function $V=\sum_{i=1}^N V_i(x_i,\, x_i^*)+V_{lz}+\tau\sum \overline d_i^\top P_i\overline d_i$, where $V_{lz}\triangleq\frac{1}{2}(\lm_i-\tilde \lm_i)^\top(\lm_i-\tilde \lm_i)+\frac{1}{2}(z_i-\tilde z_i)^\top(z_i-\tilde z_i)$ and the constant $\tau>0$ will be selected later.
	
	It can be verified that item i) in \eqref{def:condition} holds. To confirm item ii),  we take the derivative of $V$ along the trajectory of \eqref{sys:agent} and \eqref{ctrl:robust}:
	\begin{align*}
	\dot{V}\leq &\sum_{i=1}^N (y_i-y_i^*)^\top (u_i-u_i^*)+ \sum_{i=1}^N (\lm_i-\tilde \lm_i)^\top\dot{\lm}_i\\
	&+ \sum_{i=1}^N (z_i-\tilde z_i)^\top\dot{z}_i+2\tau\sum_{i=1}^N\overline d_i^\top P_i\dot{\overline d}_i \\
	=&\sum_{i=1}^N (y_i-y_i^*)^\top (\hat u_i+\lm_i-u_i^*)+ \sum_{i=1}^N (\lm_i-\tilde \lm_i)^\top\dot{\lm}_i\\
	&+ \sum_{i=1}^N (z_i-\tilde z_i)^\top\dot{z}_i+2\tau\sum_{i=1}^N\overline d_i^\top P_i\dot{\overline d}_i\\
	\leq& \sum_{i=1}^N (y_i-y_i^*)^\top (\hat u_i-\hat u_i^*)+ \sum_{i=1}^N (y_i-y_i^*)^\top (\lm_i- \tilde \lm_i)+ \\
	&\sum_{i=1}^N (\lm_i-\tilde \lm_i)^\top\dot{\lm}_i+ \sum_{i=1}^N (z_i-\tilde z_i)^\top\dot{z}_i-\tau\sum_{i=1}^N\overline d_i^\top \overline d_i\\
	\leq & (y-y^*)^\top (\hat u-\hat u^*)-(\lm-\tilde \lm)^\top L (\lm-\tilde \lm)\\
	&-\sum_{i=1}^N (\lm_i-\tilde \lm_i) D_i^\epsilon \overline d_i-\tau\sum_{i=1}^N\overline d_i^\top \overline d_i
	\end{align*}
	where $\lm=\mbox{col}(\lm_1,\,\dots,\,\lm_N)$,\, $\tilde\lm=\mbox{col}(\tilde\lm_1,\,\dots,\,\tilde\lm_N)$ and $\hat u_i(y_i)=-\gamma \nabla f_i(y_i)+{\bf u}_i(y_i)-{\bf u}_i(y_i^*)$. 
	
	From Assumption \ref{ass:graph}, $(\lm-\tilde \lm)^\top L (\lm-\tilde \lm)\geq c(\lm-\tilde \lm)^\top (\lm-\tilde \lm)$ where $c$ is the minimal positive eigenvalue of $L$.  Using Young's inequality to $(\lm_i-\tilde \lm_i) D_i^\epsilon \overline d_i$ gives
	\begin{align*}
	\dot{V}\leq &(y-y^*)^\top (\hat u-\hat u^*)-c(\lm-\tilde \lm)^\top (\lm-\tilde \lm)\\
	&+\frac{c}{2}(\lm-\tilde \lm)^\top (\lm-\tilde \lm)+\frac{1}{2c}\sum_{i=1}^N||D_i^\epsilon||^2 ||\overline d_i||^2-\tau\sum_{i=1}^N\overline d_i^\top \overline d_i\\
	\leq&(y-y^*)^\top (\hat u-\hat u^*)-\frac{c}{2}(\lm-\tilde \lm)^\top (\lm-\tilde \lm)-(\tau-\tau^*)||\overline d||^2
	\end{align*}
	where $\tau^*=\frac{1}{2c} \max_i||D_i^\epsilon ||^2$.  Letting  $\tau\geq 1+\tau^*$ gives
	\begin{align*}
	\dot{V}&\leq -(y-y^*)^\top (\hat u-\hat u^*)-\frac{c}{2}(\lm-\tilde\lm)^\top (\lm-\tilde\lm)-||\overline d||^2
	\end{align*}
	which implies the composite system is passive w.r.t. its equilibrium.
	
	Having the passivity of \eqref{sys:agent} with output $y$ and input $\hat u$, we then prove the convergence of $y$ w.r.t. $y^*$. For this purpose, we only have to guarantee the strict passivity of $\hat u$ w.r.t. $y^*$ by Lemma \ref{lem:monotone}.  In fact,  from the strong convexity of $f_i(\cdot)$ and the Lipschitzness of ${\bf u}_i(\cdot)$ on the concerned set, we have
	\begin{align}
	(y_i-y_i^*)^\top [\hat u_i(y_i)-\hat u_i(y_i^*)]\leq (-\gamma \underline{h}_i+M_i) ||y_i-y_i^*||^2.
	\end{align}
	Taking $\gamma >\max_{i}(\frac{1+M_i}{\underline{h}_i})$ gives $(y_i-y_i^*)^\top [\hat u_i(y_i)-\hat u_i(y_i^*)]\leq -||y_i-y_i^*||^2$, which implies the strict passivity of ${\bf u}_i(\cdot)$ w.r.t. $y^*$. By Lemma \ref{lem:monotone}, it follows $\lim_{t\to +\infty} y_i= y^*_i$ for $i=1,\,\ldots,\, N$.
	
	To prove the asymptotic stability, we can check that $(x^*, \,\tilde \lm,\, z^*,\, {\bf 0})$ is the only trajectory contained in the set $\left\{(x,\,\lm,\, z,\overline d)\mid \dot{V}=0\right\}$ by the $y_i^*$-observability of agent $i$. According to LaSalle's invariance principle (\cite{khalil2002nonlinear}), one can obtain the conclusions.
\pe 

	Notably, the presented passivity-based approach provides a new control perspective for existing distributed optimization problems. Unlike the problems considered in \cite{trip2016internal,stegink2017unifying}, we aim to achieve a distributed output optimization while the optimization part happens on the input side in their formulations. Since passivity has been widely used in many nonlinear control publications (\cite{khalil2002nonlinear, wen2004unifying, pavlov2008incremental}), this method allows us considering this problem for more general physical agents other than single integrators (\cite{yi2016initialization,tang2016icca,cherukuri2015distributed}).

\begin{rem}\label{rem:integrator}
	In conventional resource allocation, the plants are actually single integrators (e.g., \cite{ cherukuri2015distributed, yi2016initialization}), which are our special cases of passivity w.r.t. $(y^*,\, x^*, \, 0)$. Thus, this conclusion is a nonlinear extension of existing results to a larger class of dynamic systems.  Furthermore, as a primal-dual based method to solve the distributed optimization problem, this algorithm is different from those in \cite{cherukuri2015distributed, kia2015distributed} which need non-trivial initializations, and this initialization-free property makes it more applicable to networked systems with variable numbers of agents.
\end{rem}

When the plants are of exponential passivity, one can further obtain the exponential convergence of this algorithm as follows.
\begin{thm}\label{thm:robust-exp}
	Under the hypothesis of Theorem \ref{thm:robust}, further assume agent $i$ is exponentially passive w.r.t. $(y^*_i,\, x^*_i,\, u^*_i)$ for $1\leq i\leq N$. Then, the distributed coordination problem with regulation constraints determined by \eqref{sys:agent} and \eqref{opt:ra} can be exponentially solved by the algorithm \eqref{ctrl:robust} with a properly chosen $\gamma$.
\end{thm}
\pb  To prove this theorem, we first let $\overline x_i=x_i-x^*_i, \overline y_i=y_i-y^*_i, \overline \lm_i=\lm_i-\tilde\lm_i,\,\overline z_i=z_i-z_i^*$, it follows
	\begin{align*}
	\dot{\overline \lm}_i&=-\sum_{j=1}^N a_{ij}(\overline \lm_i-\overline \lm_j)-\sum_{j=1}^N a_{ij}(\overline z_i-\overline z_j)+d_i^0-\overline y_i-D_i^\epsilon \overline d_i\\
	\dot{\overline d}_i&=(S_i-L_iD_i)\overline d_i\\
	\dot{\overline z}_i&=\sum_{i=j}^N a_{ij}(\overline z_i-\overline z_j).
	\end{align*}
	The whole dynamic can be written in a compact form:
	\begin{align*}
	\dot{\overline \lm}&=-L\overline \lm- L\overline z-\overline y- D^\epsilon \overline d\\
	\dot{\overline d}&=\overline S\overline d\\
	\dot{\overline z}&=L\overline \lm
	\end{align*}
	where $\overline \lm=\mbox{col}(\overline \lm_1,\, \dots,\,\overline \lm_N), \overline z=\mbox{col}(\overline z_1,\,\dots,\,\overline z_N), \overline y=\mbox{col}(\overline y_1,\,\\ \dots,\,\overline y_N),\, \overline d=\mbox{col}(\overline d_1,\,\dots,\,\overline d_N), \, \overline S=\mbox{diag}\{S_1-L_1D_1,\,\dots,\,\\S_N-L_ND_N\}$ and $D^\epsilon=\mbox{diag}\{D_1^\epsilon,\,\dots,\,D_N^\epsilon\}$. 
	
	Letting $\hat z_1=r^\top \overline z,\, \hat z_2=R^\top \overline z$ gives
	\begin{align}\label{ctrl:reduced}
	\begin{split}
	\dot{\overline \lm}&=-L\overline\lm -LR\hat z_2-\overline y -D^\epsilon \overline d\\
	\dot{\hat z}_2&=R^\top L \lm \\
	\dot{\overline d}&=\overline S\overline d
	\end{split}
	\end{align}
	where we use $r^\top \dot{\overline z}=0$ and $\hat z_1 \equiv {0}$.  Then our problem is reduced to prove the exponential stability of the composite system determined by the evolution of $\overline x,\, \overline \lm,\, \overline d,\, \hat z_2$ under \eqref{sys:agent},\,\eqref{ctrl:robust}, and \eqref{ctrl:reduced}.
	
	The proof will be accomplished by two steps.
	
	First, we prove the exponential stability of \eqref{ctrl:reduced} when $\overline y \equiv {0}$. Noticing that the system is in a cascaded form, we only have to prove the stability of the $(\overline \lm,\, \hat z_2)$-subsystem when $\overline y \equiv {0}$ and  $\overline d  \equiv0$, since $\overline S$ is already Hurwitz by the selection of $L_i$.
	
	Since system \eqref{ctrl:reduced} is linear, we only have to obtain the asymptotic stability of $(\overline \lm,\, \hat z_2)$-subsystem. For this purpose, we consider  $V_{lz}=\frac{1}{2}\overline \lm^\top \overline\lm+\frac{1}{2}\overline z_2^\top \overline z_2$ as a Lyapunov candidate, and then its derivative along the trajectory of $\eqref{ctrl:reduced}$ when $\overline y \equiv {0}$ and  $\overline d  \equiv0$ satisfies the following:
	\begin{align*}
	\dot{V}_{lz}&\leq -\overline\lm^\top L\overline\lm\leq - c  \overline\lm^\top  \overline\lm
	\end{align*}
	where $c$ is the minimal positive eigenvalue of $L$. This consequently implies $\overline \lm\to 0$ as $t$ goes to infinity. Denote $\overline A\triangleq \begin{bmatrix}
	-L&LR\\
	R^\top L& 0
	\end{bmatrix}$. Under Assumption \ref{ass:graph}, $LR$ has a full column-rank, then the pair $\left([I~~0],\, \overline A \right)$ is observable by the PBH-test. Combining the above arguments, we can conclude the asymptotic stability of $(\overline \lm,\, \hat z_2)$-subsystem when $\overline y \equiv {0}$ and  $\overline d  \equiv0$ and thus the exponential stability of $\eqref{ctrl:reduced}$
	when $\overline y \equiv {0}$.
	
	Next, we prove the exponential stability of composite system. Since $\eqref{ctrl:reduced}$ is exponential stable when $\overline y \equiv {0}$ , there exists a unique positive definite matrix $P$ satisfying  $\hat A^\top P+P\hat A=-I$ for $\hat A\triangleq\mbox{diag}\{\overline A,\,\overline S\}$. Take a Lyapunov candidate for the composite system as $\overline V=\sum_{i=1}^N V_i(x_i,\, x_i^*)+\hat \lm^\top P \hat \lm$ with $\hat \lm=\mbox{col}(\overline \lm,\, \hat z_2,\, \overline d)$, which is apparently positive definite due the exponential passivity of \eqref{sys:agent} by assumptions. Its derivative along the trajectory of this composite system composed of \eqref{sys:agent} and \eqref{ctrl:robust} satisfies
	\begin{align*}
	\dot{\overline V}&\leq -\sum_{i=1}^N c_{i1}V_i-\sum_{i=1}^N\overline y_i^\top [\hat u_i(y_i)-\hat u_i(y_i^*)]+\overline y^\top \overline\lm \\
	&-\hat \lm^\top \hat \lm+2\hat \lm^\top P\hat A \hat B \overline y
	\end{align*}
	where $\overline B=\mbox{col}(I_N,\, {\bf 0},\, {\bf 0})$. Using Young's inequality and $||\overline\lambda||^2\leq ||\hat \lambda||^2$, one can obtain
	\begin{align*}
	\dot{\overline V}&\leq -\sum_{i=1}^N c_{i1}V_i-\sum_{i=1}^N\overline y_i^\top [\hat u_i(y_i)-\hat u_i(y_i^*)]+||\overline y||^2\\
	&+\frac{1}{4}||\overline\lm||^2 - \hat \lm^\top \hat \lm+ \frac{1}{4}||\hat \lm||^2+ 4||P\overline A \overline B||^2 ||\overline y||^2\\
	&=-\sum_{i=1}^N c_{i1}V_i- \frac{1}{2}\hat \lm^\top \hat \lm +(1+4||P\overline A \overline B||^2)||\overline y||^2\\
	&+\sum_{i=1}^N\overline y_i^\top [\hat u_i(y_i)-\hat u_i(y_i^*)].
	\end{align*}
	Taking $\gamma >\max_{i}(\frac{2+4||P\overline A \overline B||^2+M_i}{\underline{h}_i})$ gives
	\begin{align*}
	\dot{\overline V}&\leq -\sum_{i=1}^N c_{i1}V_i- \frac{1}{2}||\hat \lm||^2 - ||\overline y||^2.
	\end{align*}
	Applying Theorem 4.10 in \cite{khalil2002nonlinear} gives the exponential convergence of $\overline V$  under this algorithm and thus $y$ exponentially converges to the optimal solution of \eqref{opt:ra}. The proof is complete.
\pe

\begin{rem}
	In contrast to existing constrained steady-state regulation problem (\cite{jokic2009constrained}), we consider its distributed extensions where the steady-state of agents can only be determined and reached in a distributed way, which is of course more challenging. Moreover, unknown observation disturbances are taken into consideration, along with both the asymptotic and exponential convergence results, while only local and asymptotic results were obtained in \cite{jokic2009constrained, zhang2011extremum}.
\end{rem}

\begin{figure}
	\centering
	\begin{tikzpicture}[shorten >=1pt, node distance=1.2 cm, >=stealth',
	every state/.style ={circle, minimum width=0.2cm, minimum height=0.2cm}, auto]
	\node[align=center,state](node1) {1};
	\node[align=center,state](node2)[right of=node1]{2};
	\node[align=center,state](node3)[right of=node2]{3};
	\node[align=center,state](node4)[right of=node3]{4};
	\path[-]   (node1) edge (node2)
	(node2) edge [bend left] (node4)
	(node2) edge [bend right] (node3)
	(node3) edge [bend right]  (node4)
	;
	\end{tikzpicture}
	\caption{The communication graph $\mathcal{G}$.}\label{fig:graph}
\end{figure}
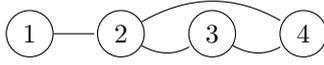

\section{Applications and Discussions}\label{sec:discussion}
In this section, we provide applications of previous designs and examples to verify the effectiveness.

\subsection{Distributed Inventory Control}
In this subsection, we show how a distributed inventory control problem can be formulated as a resource allocation problem over dynamic agents and solved by our approach.  We consider only one perishable commodity and $N$ networked inventories (\cite{raafat1991survey}). 

The inventory system at node $i$ is modeled as
\begin{align}\label{sys:inventory}
\dot{I}_i=-\theta_i I_i+P_i-D_i,
\end{align}
where $I_i$ is the inventory level, $\theta_i>0$ is the deterioration rate, $P_i$ is the production rate at node $i$, and $D_i$ is a constant demand rate. The information structure among these inventories is represented by a connected graph $\mathcal{G}$. The storage cost at each warehouse is given as $f_i(I_i)=\alpha_i I_i^2+\beta_iI_i+\gamma_i$, where $\alpha_i>0$.

Generally speaking, we aim to maintain the total inventory at certain level $I^{r}$ to satisfy the customer's demands and some safety goals. Thus, this inventory control problem can be formulated as follows. Given inventory systems and cost functions $f_1(\cdot),\dots,\,f_N(\cdot)$, find a production rate for each inventory in a distributed way, such that the inventory level $I$ converges to the optimal solution $I^*\triangleq \mbox{col}(I_1^*,\,\dots,\,I_N^*)$ that solves:
\begin{align}\label{opt:inventory}
\begin{split}
\mbox{minimize}&\quad  \quad \sum\nolimits_{i=1}^Nf_i(I_i)\\
\mbox{subject to}&\quad   ~~\sum\nolimits_{i=1}^N I_i=I^r.
\end{split}
\end{align}

\begin{figure}
	\centering
	\includegraphics[width=0.80\textwidth]{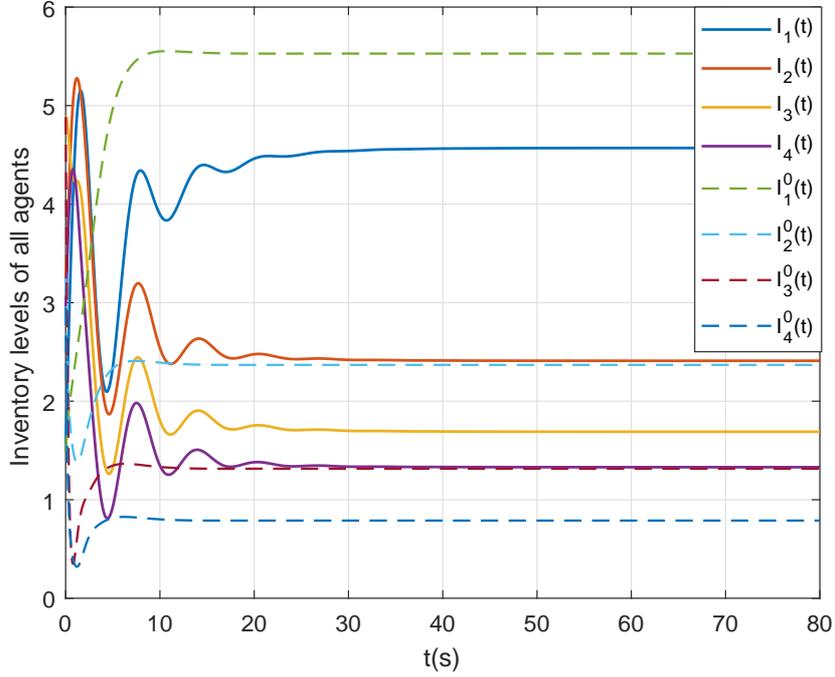}
	\caption{Profiles of the inventory levels under the control \eqref{ctrl:inventory}.}\label{fig:inventory}	
\end{figure}

Clearly, the $i$-th inventory system is exponentially passive w.r.t. $(I_i^*,\,{\bf u}_i(I_i^*))$ with input $y=I_i$, input $P_i$ and ${\bf u}_i(I_i^*)=\theta_iI_i^*+D_i$, and hence Assumptions \ref{ass:function}--\ref{ass:passivity} are also satisfied. With a pre-allocation of inventory level $I^r=\sum\nolimits_{i=1}^N I_i^r$, the following corollary shows the effectiveness of our previous design on distributed inventory control.

\begin{cor}\label{cor:inventory}
	Given the communication graph $\mathcal{G}$ and cost functions $f_1(\cdot),\,\dots,\, f_N(\cdot)$, the distributed inventory control problem determined by \eqref{sys:inventory} and \eqref{opt:inventory} can be solved by the following algorithm
	\begin{align}\label{ctrl:inventory}
	\begin{split}
	P_i&=-\nabla f_i(I_i)+\lm_i+{\bf u}_i(I_i)\\
	\dot{\lm}_i&=-\lm^v_i-z^v_i+I_i^r-I_i\\
	\dot{z}_i&=\lm^v_i,\quad i=1,\,\dots,\,N
	\end{split}
	\end{align}
	where $\gamma$ can be any positive constant. Moreover, $I(t)$ converges to $I^*$ exponentially as $t\to \infty$.
\end{cor}
	
	We then provide a numerical example with four inventories having parameters $\alpha_i=0.1 i,\, \beta_i=-0.05 i,\,  \gamma_i=\theta_i=D_i=I_i^r=i$, $i=1,\,\dots,\, 4$. The communication graph is taken as Fig.~\ref{fig:graph} and all initial conditions are randomly chosen in $[0,\,6]$. By choosing control inputs as \eqref{ctrl:inventory}, we solve this inventory control problem and drive the outputs to the optimal solution $I^*=\mbox{col}(4.57,\, 2.41,\,1.69,\, 1.33)$. For comparisons, we take the input $P_i$ as that in \cite{yi2016initialization}, and the output trajectories of inventories are represented by dash lines. It can be found the algorithm fails to solve our problem and only drives the outputs of agents to a non-optimal point $\mbox{col}(5.53,\,2.37,\,1.32,\,0.79)$, which confirms the effectiveness of our design.
	
	\subsection{Average Consensus with Disturbance Rejection}
	Consensus and especially average consensus of multi-agent agents has been shown as an inevitable part of the solution for more complex problems in several applications, including distributed filtering and multi-robot flocking \cite{olfati2007consensus, renwei2011springer}. While consensus only requires the agreement on some common signal, an extra condition has to be satisfied in average consensus, which relates the limiting behavior of the whole system to the initial states. The average consensus problem is certainly more challenging especially when we expect an average consensus of all outputs for a heterogeneous multi-agent network. 

\begin{figure}
	\centering
	\includegraphics[width=0.80\textwidth]{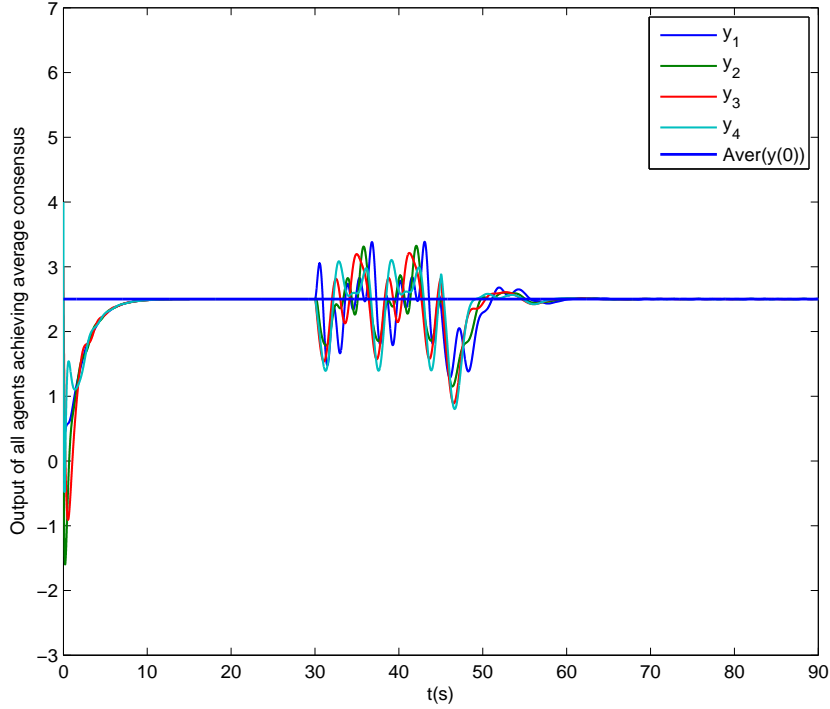}
	\caption{Profiles of the outputs achieving average consensus. }\label{fig:aver}
\end{figure}

In our formulation, let $f_i(s)=\frac{1}{2}s^2$ and one can obtain the following conclusion.
\begin{cor}\label{cor:equal-ra}
	Under Assumptions \ref{ass:graph}--\ref{ass:passivity}, the outputs of agents \eqref{sys:agent} can reach the average of their private data $d_i^0$ under the following algorithm
	\begin{align}\label{ctrl:equal-ra}
	\begin{split}
	u_i&=-\gamma y_i+\lm_i+{\bf u}_i(y_i)\\
	\dot{\lm}_i&=-\lm^v_i-z^v_i+d_i-y_i-D_i^\epsilon \eta_i\\
	\dot{\eta}_i&=(S_i-L_iD_i)\eta_i+L_id_i\\
	\dot{z}_i&=\lm^v_i,\quad i=1,\,\dots,\,N
	\end{split}
	\end{align}
	where $\gamma > 1+\max_{i}M_i$ and $L_i$ is selected as in Theorem \ref{thm:robust}, i.e., $\lim_{t\to +\infty} y_i= \frac{1}{N}\sum_{i=1}^Nd_i^0$ for $i=1,\,\ldots,\,N$.
\end{cor}
The proof is a direct application of Theorem \ref{thm:robust}. As a byproduct of this corollary, we can solve the output average consensus problem of these agents in spite of observation disturbances $d_i^\epsilon$ by letting $d_i^0=y_i(0)$ . Then, $y_i\to {\rm Aver}(y(0))\triangleq \frac{1}{N}\sum_{i=1}^N y_i(0)$ as $t$ goes to infinity. Since single integrator is passive, this result extends the average consensus results to a larger class of nonlinear systems with disturbance rejection. 

To verify the effectiveness of this algorithm, we consider four controlled Chua's circuits (\cite{lu2007impulsive}) as follows.
\begin{align*}
\dot{x}_{i1}&=\alpha_i(x_{i2}-x_{i1}-f_i(x_{i1})+F_i)\\
\dot{x}_{i2}&=x_{i1}-x_{i2}+x_{i3}\\
\dot{x}_{i3}&=-\beta_i x_{i2}\\
y_i&=x_{i1}
\end{align*}
where $F_i$ is the input signal and $f_i(x_{i1})=b_ix_{i1}+\frac{1}{2}(a_i-b_i)(|x_{i1}+c_i|-|x_{i1}-c_i|)$ with typical parameters $\alpha_i=9,\,\beta_i=\frac{100}{7},\,a_i=-\frac{8}{7},\,b_i=-\frac{5}{7},\, c_i=1$, The agents are coupled by a communication graph as Fig.~\ref{fig:graph}.  We aim to achieve an output average consensus by output feedback control.  
	
First, we let $F_i=u_i+f_i(x_{i1})$ to passivate agent $i$'s dynamics with output $y_i$ and a new input $u_i$. In fact, Assumptions \ref{ass:regulator} and \ref{ass:passivity} hold with ${\bf x}_{i1}(r)=r$,\,${\bf x}_{i2}(r)=0$,\,${\bf x}_{i3}(r)=-r$, ${\bf u}_i(r)=r$, and $V_i(x,\, x^*_i)= \frac{1}{\alpha_i}(x_{i1}-x_{i1}^*)^2+(x_{i2}-x_{i2}^*)^2+ \frac{1}{\beta_i}(x_{i3}-x_{i3}^*)$. Thus the average output consensus problem among these agents can be solved by \eqref{ctrl:equal-ra}.
	
For simulations, we assume agent $i$ is subject to a sinusoidal observation disturbance with frequency $\w_i=4-i$ and unknown amplitude or phase after $t=30\,{\rm s}$. The controller is taken as \eqref{ctrl:robust} with $D_i^\epsilon=[0,\,0]$ during $0\,{\rm s}\sim45\,{\rm s}$ and with $D_i^\epsilon=[1,\,0]$ after $45\,{\rm s}$. All initial conditions are randomly chosen in $[-5,\,5]$. At first, the outputs of agents quickly converge to their average point. Then, the average consensus is disrupted by those observation disturbances. After the disturbance rejection part works at $t=45\,{\rm s}$, we recover the output average consensus of these agents.  The detailed performance of the above control is depicted in Fig.~\ref{fig:aver}.

\subsection{Non-minimum Phase Multi-Agent Coordination}
	
	Note that many non-minimum phase nonlinear systems have an incremental passivity property (perhaps after a passivation procedure) \cite{khalil2002nonlinear}. It is appealing to employ the proposed algorithms to handle non-minimum phase nonlinear agents with more complicated objective functions. We present an example in this subsection to verify this point.
	
	Consider a network of four nonlinear agents described by 
	\begin{align*}
	\dot{z}_{i1}&=\e_{i1}z_{i2}^3,\\
	\dot{z}_{i2}&=-\e_{i2}z_{i1}+\e_{i3}x_i,\\
	\dot{x}_i&=-\e_{i4}z_{i2}^3-{\e_{i5}}x_i+u_i,\\
	y_i&=x_i
	\end{align*}
	where $\e_{i1}, \dots, \e_{i5}$ are positive constants ($i=1,\,\dots,\,4$). 
	
	Apparently, the zero dynamics of agent $i$ is $\dot{z}_{i1}=\e_{i1}z_{i2}^3,\,\dot{z}_{i2}=-\e_{i2}z_{i1}$, which is not asymptotically stable. Thus, the agents are all non-minimum phase.  Nevertheless, it can be verified that Assumptions \ref{ass:regulator} and \ref{ass:passivity} hold with ${\bf z}_{i1}(r)=\frac{\e_3}{\e_2}r, {\bf z}_{i2}(r)=0, {\bf x}_i(r)=r, \, {\bf u}_i(r)=\e_{i5}r$ and storage functions 
	\begin{align*}
	V_i= \frac{1}{2}z_{i1}^2+ \frac{\e_{i1}}{4\e_{i3}}z_{i2}^4+ \frac{\e_{i1} \e_{i3}}{2\e_{i2}\e_{i4}} x^2_i.
	\end{align*}
	
Take $\e_{ij}=1$ and an information sharing graph as in Fig.~\ref{fig:graph}. We consider the distributed coordination problem among these agents with regulation constraints. The local cost functions satisfying Assumption \ref{ass:function} are chosen as $f_1(y_1)=(y_1+3)^2,\,f_2(y_2) = y_2^2 \ln(1 + y_2^2) + (y_2+1)^2$, $f_3(y_3)=\ln(e^{-0.1y_3}+e^{0.3y_3})+y_3^2$ and $f_4(y_4)=\frac{y_4^2}{25\sqrt{y_4^2 + 1}}+(y_4-3)^2$. Assume the constant $d_i^0=i$ and agent $i$ is subject to a sinusoidal disturbance with frequency $\w_i=4-i$ but unknown amplitude or phase after $t=75\,{\rm s}$. The problem is solvable by Theorem \ref{thm:robust}.
	
\begin{figure}
	\centering
	\includegraphics[width=0.80\textwidth]{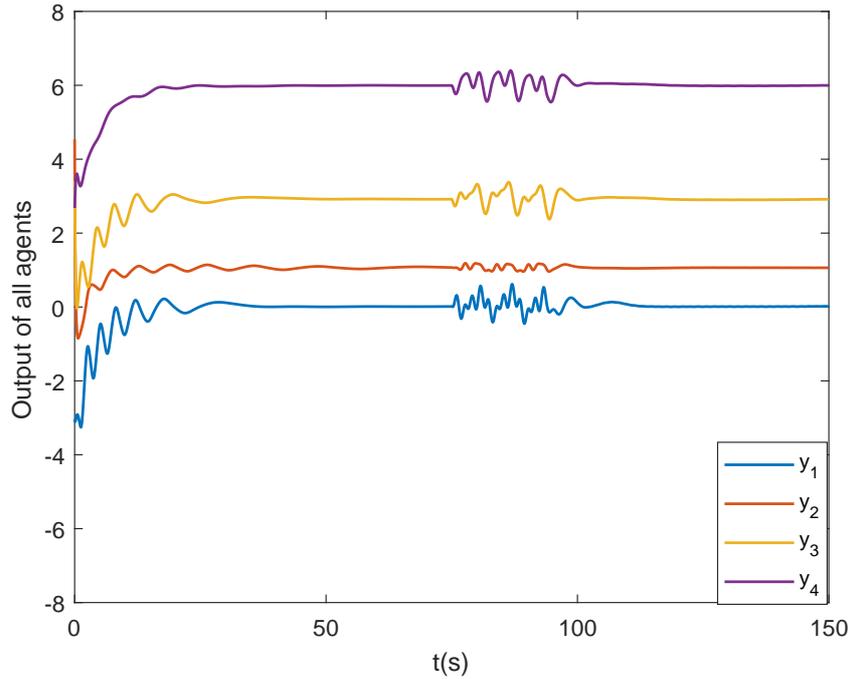}
	\caption{Profiles of the outputs $y_i$ under our controller \eqref{ctrl:robust}.}\label{fig:simu}
\end{figure}

Choose $\gamma=2$, $L_1=\mbox{col}(5.00,\,6.72,\,2.19)$,\,$L_2=\mbox{col}(5.00,\,6.51,\,2.75)$,\, $L_3=\mbox{col}(5.00,\,6.07,\,3.69)$, and $L_4=\mbox{col}(5.00,\,5.00,\,5.00)$.  To verify the disturbance rejection performance, we let $D_i^\epsilon=[0,\,0]$ during $0\,{\rm s}\sim95\,{\rm s}$ and $D_i^\epsilon=[1,\,0]$ after $95\,{\rm s}$. The evolution of $y_i$ under \eqref{ctrl:robust} is depicted in Fig.~\ref{fig:simu}.  At first, all outputs of agents evolve without disturbances and quickly converge the optimal point. Then, the agents are moved away from the optimal steady-state due to the observation disturbances. After the disturbance rejection part works at $t=95\,{\rm s}$, we recover the optimal steady-state regulation of these agents, which confirms the conclusions.

\section{Conclusions}\label{sec:con}
A distributed coordination problem with regulation constraints was formulated and solved for a class of nonlinear passive multi-agent systems in this paper. By reviewing the passivity technique with respect to non-zero equilibria, we reduce the concerned optimization to a passivity-based regulation problem. Combined with graph theory and observer design technique, gradient-based rules are proposed to solve our problem with disturbance rejection. Potential applications and numerical examples were presented to show their effectiveness. In fact, many interesting and challenging problems still remain to be addressed, including how to solve this problem  under switching graphs and extend the gradient-based rules to general monotone-operator-based designs.

\bibliographystyle{iet}
\bibliography{opt-passivity-j}
\nocite{*}

\end{document}